# Algorithm for constructing customized quantized resistances in graphene *p-n* junctions

Albert F. Rigosi[1*], Martina Marzano[2,3], Antonio Levy[1], Heather M. Hill[1], Dinesh K. Patel[1,4], Mattias Kruskopf[1,5], Hanbyul Jin[1,5], Randolph E. Elmquist[1], and David B. Newell[1]

*[1]Physical Measurement Laboratory, National Institute of Standards and Technology (NIST), Gaithersburg, MD 20899, United States*

*[2]Department of Electronics and Telecommunications, Politecnico di Torino, Torino 10129, Italy*

*[3]Istituto Nazionale di Ricerca Metrologica, Torino 10135, Italy*

*[4]Department of Physics, National Taiwan University, Taipei 10617, Taiwan*

*[5]Joint Quantum Institute, University of Maryland, College Park, MD 20742, United States*

ABSTRACT: A mathematical approach is introduced for predicting quantized resistances in graphene *p-n* junction devices that utilize more than a single entry and exit point for electron flow. Depending on the configuration of an arbitrary number of terminals, electrical measurements yield nonconventional, fractional multiples of the typical quantized Hall resistance at the $\nu = 2$ plateau ($R_\mathrm{H} \approx 12906\ \Omega$) and take the form: $\frac{a}{b} R_\mathrm{H}$. This theoretical formulation is independent of material, and applications to other material systems that exhibit quantum Hall behaviors are to be expected. Furthermore, this formulation is supported with experimental data from graphene devices with multiple source and drain terminals.

---

[*] Email: afr1@nist.gov

# 1. INTRODUCTION

Graphene has been studied extensively over the past decade due to its exceptional optical and electrical properties [1-4], and devices made from this material display quantized Hall resistance values of $\frac{1}{(4m+2)}\frac{h}{e^2}$, where $m$ is an integer, $h$ is the Planck constant, and $e$ is the elementary charge. When configured as *p-n* junctions (*pn*Js) [5-18], specifically with the intent of exploring transport in the quantum Hall effect (QHE), these graphene-based devices become a foundation for exploring two-dimensional physics. Concordantly, when properly and fully understood, such devices can be applied towards photodetection [19-23], electron optics [24-27], and quantum Hall resistance standards [28-38].

Of the applications listed, the latter has benefitted from the prospect of using *pn*Js for quantum transport to access different quantized values of resistance [39-41]. The behavior of those *pn*Js depend heavily on which Landauer-Büttiker edge states equilibrate at the relevant junctions and an extensive analysis of these behaviors was explained in several reports, some involving tunable gates used to adjust the *pn*J [5-8, 42-46]. To date, most of these studies have only used a single source and drain, and though this has been useful for many fundamental device measurements, this limit on how the current is permitted to flow confines the parameter space within which a device may be operated.

This work introduces a new mathematical approach to predicting quantized resistances in *pn*J devices utilizing more than a single entry and exit point for electron flow. The introduction of multiple electron entry or exit terminals appears to contribute to a nonintuitive effect on the measurable effective resistance of the device. For different numbers and configurations of terminals, electrical measurements yield a variety of nonconventional, fractional multiples of the typical quantized Hall resistance at the $\nu = 2$ plateau ($R_{\text{H}} \approx 12906\ \Omega$) and take the form: $\frac{a}{b} R_{\text{H}}$, where $a$ and $b$ are integers. As seen in a previous experimental work [49], $a$ and $b$ formed several measurable coefficients of $R_{\text{H}}$. Since those observations, an explanation for why those values appeared was not fully understood, let alone formulated into a theoretical framework. The latter framework is important to understand and may provide insights on device functionality in the quantum Hall regime. Furthermore, this new framework provides an easily implementable algorithm for the calculation of the quantized resistances when specific experimental outputs are desired, an approach that cannot be granted by simulations alone. Given that this formulation is independent of material properties, applications to other material systems that exhibit quantum Hall behavior are to be expected.

## 2. NUMERICAL AND EXPERIMENTAL METHODS

### 2.1. Simulations for *p-n* Junction Devices with Multiple Terminals

The use of the terminology 'source' and 'drain' will not be used to prevent confusion regarding electron flow within the device and will be replaced by electron entry or exit terminals, with entry and exit terminals being representing by negative and positive symbol icons, respectively. Many simulations were performed to determine a generalized behavior of how the overall effective resistance of the circuit shown in Fig. 1 changes with terminal configuration. These simulations were performed with the analog electronic circuit simulator LTspice [47-50].

The overall circuit comprises quantized regions of both *p*-type and *n*-type variety, with both regions being modeled as either ideal counterclockwise (CCW) or clockwise (CW) *k*-terminal quantum Hall effect (QHE) elements and shown as gray or blue regions in Fig. 1, respectively. The terminal voltages $e_m$ are related to the currents $j_m$ by $R_{\text{H}} j_m = e_m - e_{m-1}$ ($m = 1, \ldots, k$) for CW elements and $R_H j_m = e_m - e_{m+1}$ for CCW elements. To determine the circuit's behavior at the external terminals (labeled as $A$ and $B$ in Fig. 1), only a single magnetic field ($B$-field) direction was simulated at a time. For a positive $B$-field, an $n$-type ($p$-type) device (experimentally represented as a graphene-based device) was modeled by a CW (CCW) QHE element, whereas, when $B$ is negative, a CWW (CW) QHE element was used. Concrete examples are provided and fully explained in the Supplemental Material [51].

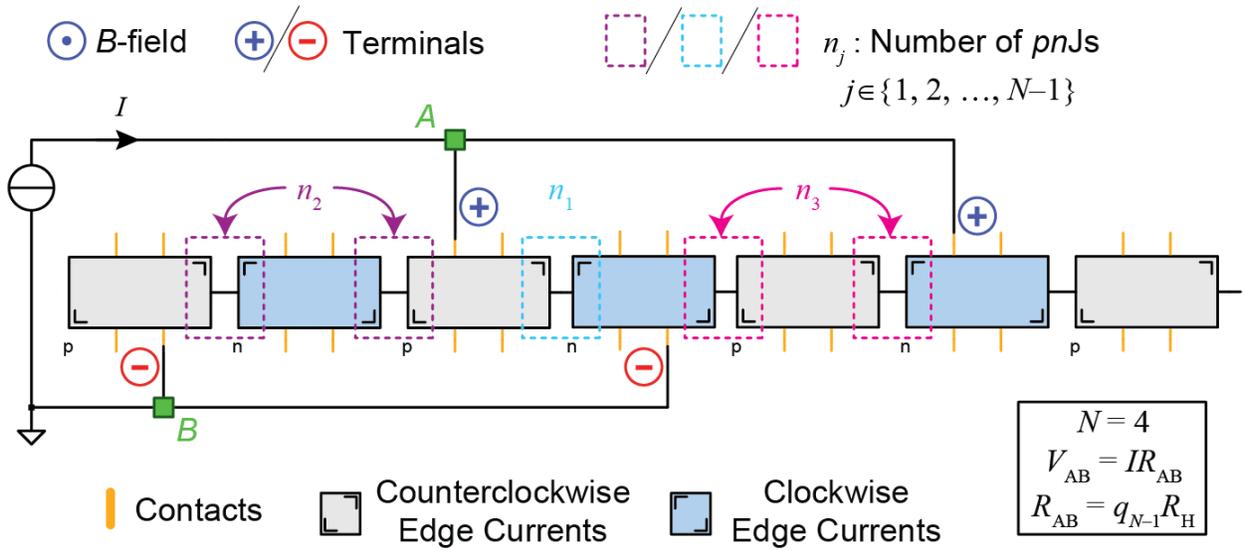

FIG. 1. (Color online) A $pn$J device is illustrated as represented by the schematic drawn in the LTspice circuit simulator. Gray and blue regions define the counterclockwise (CCW) and clockwise (CW) edge currents, as predicted in a positive magnetic field for $p$-type and $n$-type doping, respectively. There is a total of 4 entry/exit terminals in this depiction (two entries and two exits, totaling $N$), with the various $n_j$ values labeled in purple, light blue, and pink dashed rectangles. The effective resistance of the device is found from its corresponding voltage measurement between points $A$ and $B$ (green squares). The assignment of numerical subscripts for the various opposing polarities of terminals is described later in the main text.

## 2.2. Device Fabrication

To test some of the simulations, graphene *pn*J devices were fabricated from epitaxially grown graphene (EG). When at the high temperature of 1900 °C, Si atoms sublimate from SiC samples, leaving behind an excess of carbon that arranges itself into a honeycomb lattice. Samples were diced from on-axis 4*H*-SiC(0001) semi-insulating wafers (CREE) [50]. After a chemical cleaning with a 5:1 diluted solution of hydrofluoric acid and deionized water, the samples were processed with AZ5214E for polymer-assisted sublimation [52]. Next, samples were placed silicon-face down on a polished graphite substrate (SPI Glas 22) [50]. The growth was performed in an ambient argon environment at 1900 °C with a graphite-lined resistive-element furnace (Materials Research Furnaces Inc.) [50], with heating and cooling rates of about 1.5 °C/s.

After growth, the quality of the EG was assessed with confocal laser scanning and optical microscopy for efficiency [53-54]. Protective layers of Pd and Au were deposited on the EG to prevent organic contamination during the subsequent photolithography processes, all of which are described in a previous work [49]. The required functionalization with $Cr(CO)_3$ to reduce the electron density to the order of $10^{10}$ cm$^{-2}$ [55-58], as well as the deposition of a S1813 photoresist spacer layer and photoactive ZEP520A layer, was also followed as previously described [49, 59].

## 2.3. Definitions for Theoretical Framework

As the theoretical framework for determining the effective quantized resistance of the circuit exemplified in Fig. 1 is explained, certain parameters will be frequently mentioned. The total number of terminals, regardless of polarity and position, is defined as *N*. For instance, the device in Fig. 1 has a total of 4 terminals. Direct current measurements were set up such that a voltage was always measured between points *A* and *B* in Fig. 1, yielding a resistance of the form $R_{AB} =$ $q_{N-1}R_H$, where $R_H$ is the Hall resistance at the $\nu = 2$ plateau ($R_H \approx 12906\ \Omega$) and $q_{N-1}$ is defined as the *coefficient of effective resistance* (CER). The CER, based on the efforts reported here, appears to be restricted to the set of rational numbers, with all simulated results being expressed as a either a fraction or an integer.

For $N$ terminals, the number $n_j$ is defined as the number of junctions between two adjacent terminals, thereby limiting $j$ to $N-1$. In the example case of Fig. 1 (where $N = 4$), there are three $n_j$ terms: $n_1 = 1$, $n_2 = 2$, $n_3 = 2$ (light blue, purple, and pink dashed rectangles in Fig. 1, respectively). Naturally, the assignment of numerical subscripts to $n_j$ must also be defined. One of the conditions upon which the forthcoming theory correctly predicts the CER is that the greater of the two outermost adjacent terminal pairs in any given configuration is assigned as $n_{N-1}$. The second outermost is then assigned as $n_{N-2}$ (if both pairs are equal, then the assignment is arbitrary and will yield the same calculated result for both selections). This alternation continues until $n_1$ has been assigned (see the example in Supplemental Material for the step-by-step exercise) [51]. Another condition for this assignment is that as one goes from region to region on the device from left to right, terminals must alternate in polarity like in Fig. 1 (negative, positive, negative, positive). More complex cases will be examined in Sec. IV.

In some sense, terminals can be treated like point charges in electrostatics (though confined to the Hall bar geometry), and this analogy of bringing a new charge from infinity to a pre-existing configuration will be useful to visualize many of the results. One assumption is that for a given terminal configuration, using either magnetic field polarity will yield the same quantized resistance. Though this was proven true for all of the performed simulations, and observed for the experimental cases [49], generalizing this observation to all possible configurations should still be treated as an assumption.

## 3. RESULTS FOR SIMPLE CONFIGURATIONS

### 3.1. The Two-Terminal Configuration

The case of two terminals is considered in Fig. 2 (a) and (b). A positive terminal is first arbitrarily placed and kept fixed while a negative terminal nearby is moved to determine $q_1$ as a function of $n_1$. Fig. 2 (a) shows the negative terminal being moved one junction at a time (from $n_1 = 1$ to $n_1 = 6$), with the corresponding result of simulation plotted in Fig. 2 (b). The simulations reveal that the CER ($q_1$) is a linear function of $n_1$ and should approach infinity as $n_1$ does. This is an intuitive result since it essentially reflects a traditional two-terminal measurement across an increasing number of Hall bar regions. Furthermore, the case of $n_1 = 0$ yields a CER of one and represents the traditional measurement across a single region. The cases where both a positive and negative terminal occupy the same region will play a more substantial role in Sec. IV.

### 3.2. The Three-Terminal Configuration

Next, the three-terminal configuration is considered in Fig. 2 (c) and (d), with each terminal alternating in polarity going from left to right. The alternating polarities will make generalizing behaviors easier for reasons that will be explained in Sec. IV. By performing simulations for $q_2$ as a function of $n_2$, the non-linear behavior begins to emerge. Two example sets of simulations are shown in Fig. 2 (c) while keeping $n_1$ fixed. In all cases of $n_1$, the resulting dependence of $q_2$ on $n_2$ is sigmoidal, warranting the use of the following ansatz:

$$q_2(x) = A_2 + \frac{A_1 - A_2}{1 + \left(\frac{x}{x_0}\right)^p}$$

(1)

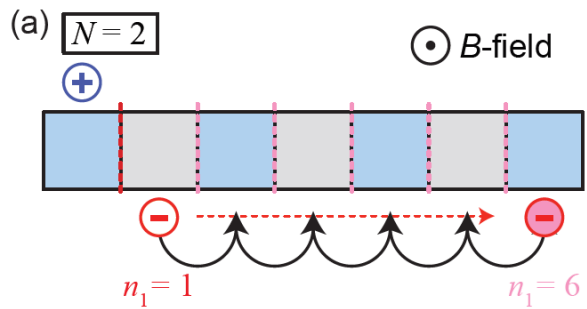

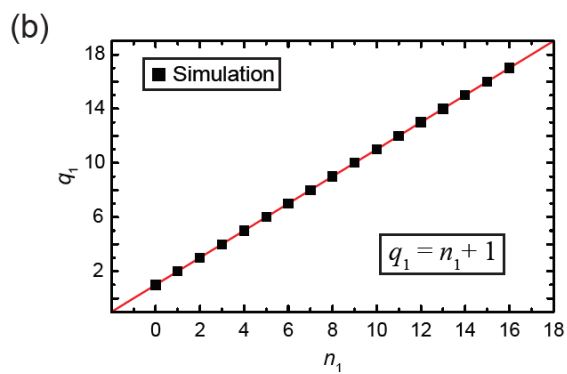

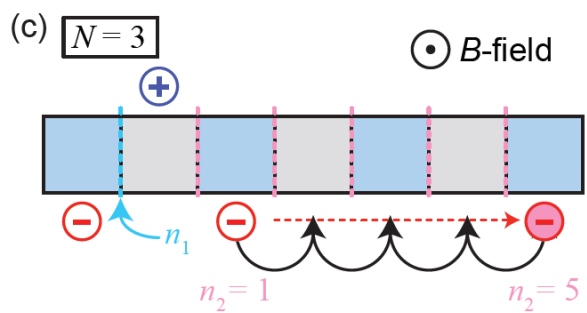

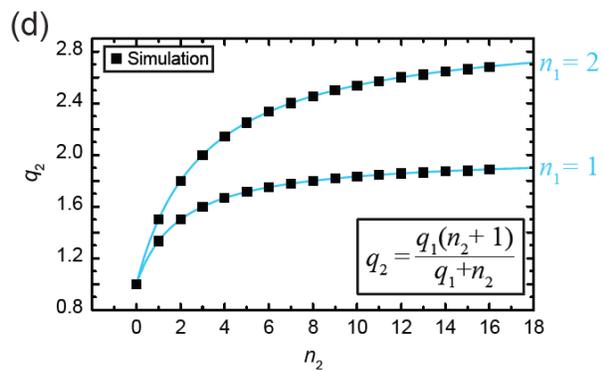

FIG. 2. (Color online) (a) An illustration of a two-terminal configuration is shown whereby the negative terminal is moved along the device. At each value of $n_1$, a simulation is performed to get the behavior of the CER ($q_1$). (b) The corresponding simulations for (a) are plotted here, yielding the intuitive result that if $n_1$ increases, $q_1$ increases linearly. (c) An illustration of a three-terminal configuration with alternating polarities is shown with fixed $n_1$ and varying $n_2$. (d) For $n_1 = 1$, one set of simulations was performed and represented by the bottom curve. The same was done for the case $n_1 = 2$. The simplified formula for the three-terminal configuration is provided in the bottom right corner.

Equation (1) is known as the Hill-Langmuir equation and is frequently seen in biochemistry and pharmacology [60] (it is also known as the Logistic fit curve in OriginLab [50]). Using this sigmoidal fit function, as opposed to others like the Fermi-Dirac distribution or the Verhulst growth model, yields zero error when fitted to the simulation data. Furthermore, the parameters of Eq. (1) take on meaningful quantities ($p = 1$):

$$q_2(n_2) = q_1 + \frac{1 - q_1}{1 + \frac{n_2}{q_1}} = \frac{q_1(n_2 + 1)}{q_1 + n_2}$$

$$(2)$$

Equation (2) establishes the general behavior for the three-terminal configuration (recall that $q_1 = n_1 + 1$). This includes configuration conditions that have not yet been fully elaborated, such as the case where two adjacent terminals are of the same polarity or if two terminals of opposite polarity occupy the same region (again, the latter case gives a well-known quantized resistance of $R_H$ and CER of $q_1 = 1$).

## 4. RESULTS FOR COMPLEX CONFIGURATIONS

### 4.1. Edge State Picture

As more terminals are added to a configuration, it becomes more helpful to have a visual representation for complex cases. For the most part, electrons flow is restricted to the regions along the entire length of the device, as defined by the endpoint terminals. For instance, in Fig. 2 (c), electrons flow from the two negative terminals to the single positive one, effectively uniting the device as a single quantized resistor with a span of $n_1 + n_2$ junctions. In these cases, the terminal configuration is said to be fully *self-interacting*. However, there are cases where configurations, like the one shown in Fig. 3 with non-alternating terminals, require a semiclassical edge state picture to assist with our visual understanding of the theoretical framework [61].

Fig. 3 (a) shows a self-interacting, four-terminal configuration whose simulations reveal that the CER holds a similar dependence on $n_x$ as it does in Fig. 2 (d) for $n_2$. As shown in Fig. 3 (a) ($n_x = 2$), the corresponding simulation yields $q_3 = \frac{12}{11}$, and the reason for using $n_x$ instead of the outermost adjacent pair will be explained later along with the formula for $N = 4$ configurations. When simulating the CER for increasing $n_x$, the sigmoidal trend only holds true for *even* values of $n_x$. For simulations of this configuration when $n_x$ is *any odd* number, $q_3 = 1$. This behavior can be explained by Fig. 3 (b), where the edge current picture provides a possible answer. Both electron paths cross in opposite directions at the same points in each junction. Furthermore, within each of the central opaque regions, the currents are expected to equilibrate since there is an appropriate, but temporary, potential difference at the $n_x$ junctions (see Supplemental Material [51]). This leads us to the justification of treating the whole configuration like a pair of isolated two-terminal configurations. And in the case of Fig. 3 (b), each two-terminal configuration has a CER of 2, thus yielding our simulation result of $q_3 = 1$. Drawing a similar diagram for other odd values will only validate this edge state visualization. The latter configuration is said to *not* be self-interacting, but rather is a simpler arrangement of two configurations in parallel.

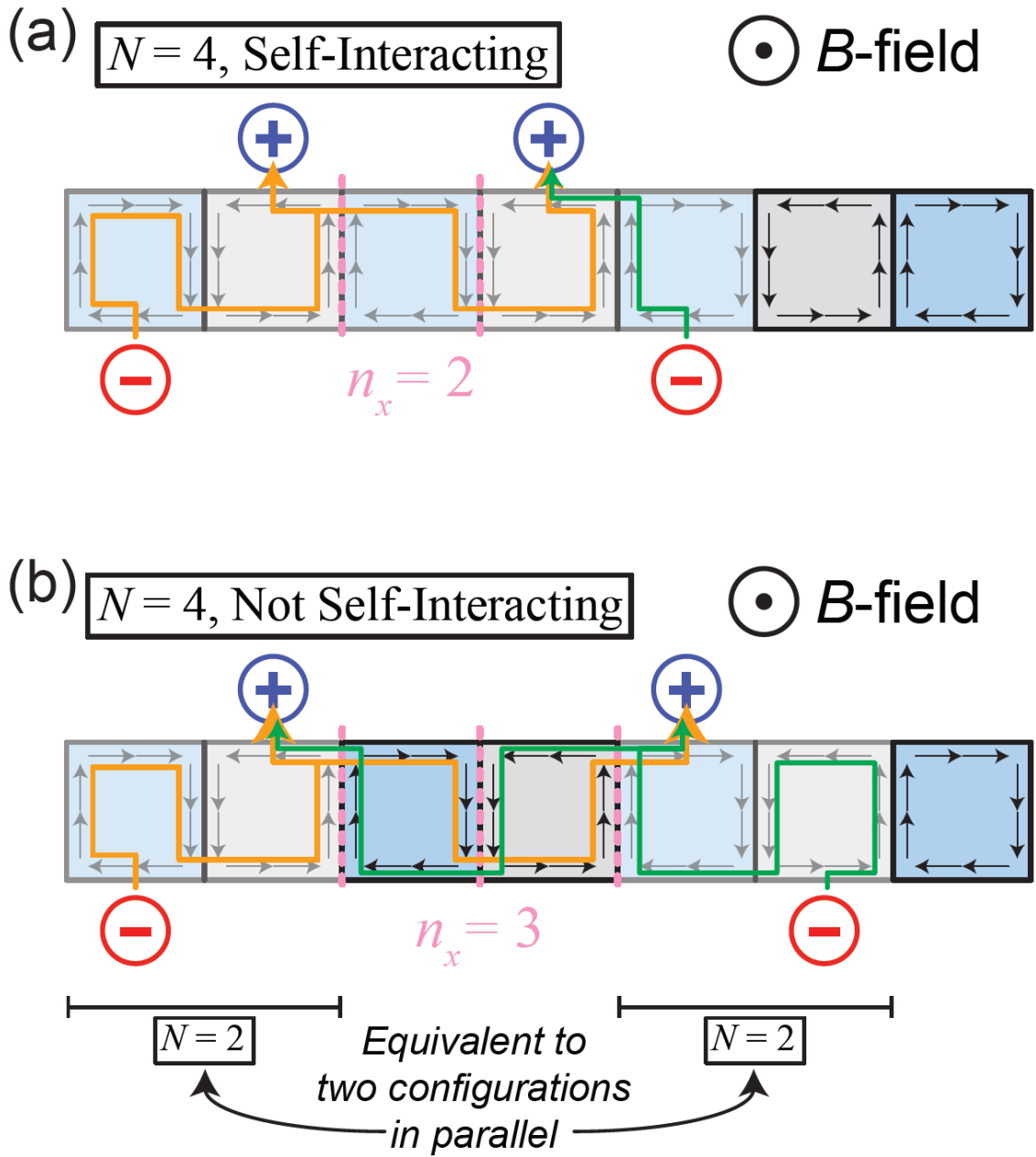

FIG. 3. (Color online) (a) An example four-terminal configuration is depicted to illustrate a visual representation of why some configurations may appear to exhibit nonintuitive quantized resistances. This subfigure is meant to be a basis of comparison for (b). (b) In the more interesting cases where a CER appears to take on a value based on a simple "parallel resistance" configuration, one can visualize the edge currents for the central, opaque regions to cancel out. This cancellation enables the CER for the two parallel $N = 2$ configurations to be calculated appropriately.

### 4.2. The Four-Terminal Configuration

When examining self-interacting, four-terminal configurations, an attempt can be made to posit a formula that describes the CER. Since it is safe to assume that configurations with alternating terminals are fully self-interacting, that is the model upon which this next analysis will depend, and the exact model is illustrated in Fig. 4 (a). As before, the two outermost adjacent pairs are inspected and the larger of the two is selected as $n_3$. The assignment of $n_2$ and $n_1$ follows thereafter in alternating succession. With all three $n_j$ terms assigned, calculating the CER becomes straightforward and starts with $q_1$. Based on the previous two formulas shown in Fig. 2, $q_1 = 3$ and $q_2 = \frac{3}{2}$. The full formula for $q_3$ has not been fully uncovered at this point, but an educated prediction can be made in the form of Eq. (2), but iteratively transformed: $q_3 = \frac{q_2(n_3+1)}{q_2+n_3}$.

This first prediction is shown in Fig. 4 (b) as a red curve, incorrectly yielding the CER of this configuration to be $q_3 = \frac{4}{3}$. Extending the simulation for $q_3$ for $n_3 : \{0, \dots, 16\}$ (black squares) shows that there is a consistent deviation from the first prediction, but not one that is solvable by a simple vertical translation. Adding a corrective term in the Hill-Langmuir equation enabled the finding of that corrective term, $\delta_3$, as a numerical value. The next question regarded the origin of the term $\delta_3$, which will be explained in Sec. IV D. For this configuration, $\delta_3 = \frac{1}{2}$ for all values of $n_3$, and the corrected formula was plotted in blue, matching the simulations with zero error:

$$q_3(n_3) = \frac{q_2(n_3 + 1)}{q_2 + n_3 + \delta_3}$$

<div align="right">(3)</div>

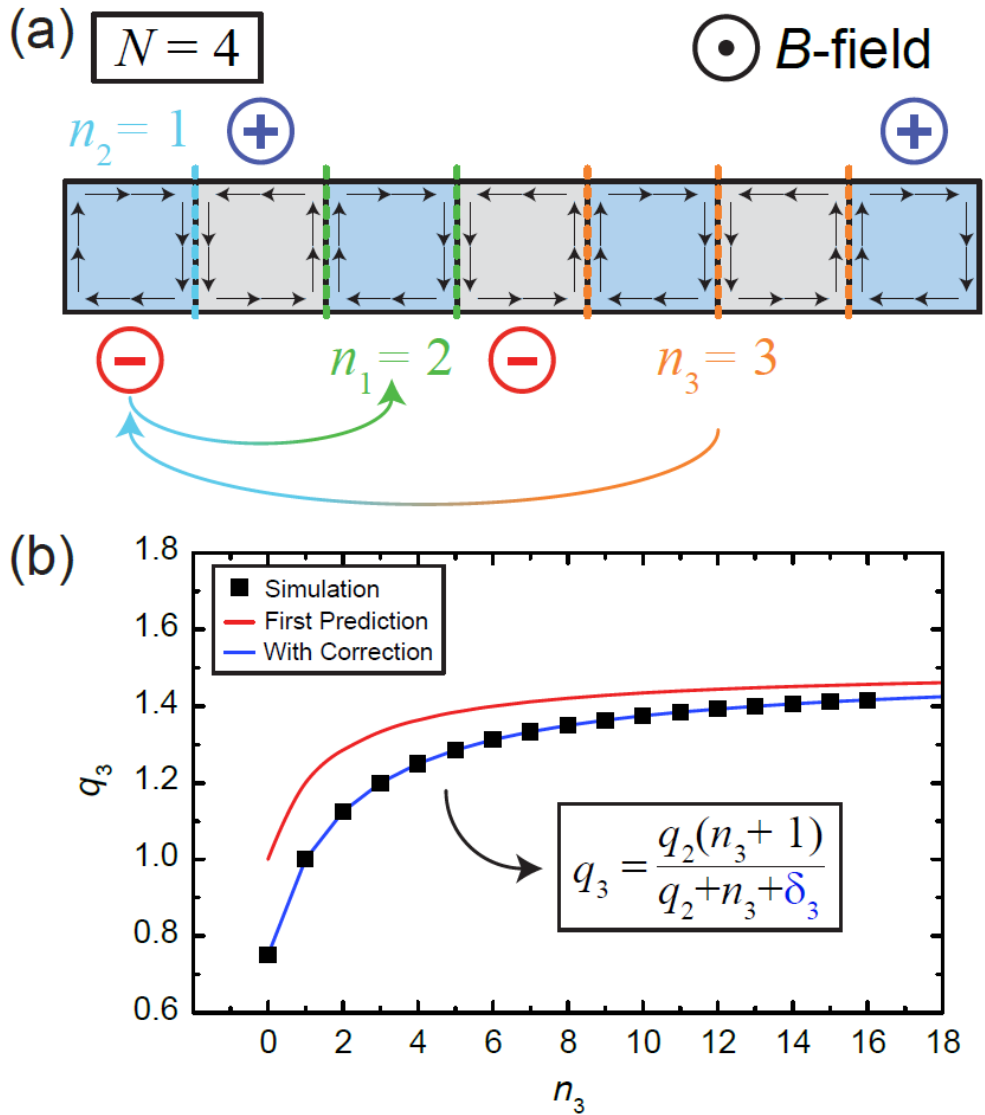

FIG. 4. (Color online) (a) A self-interacting four-terminal configuration is illustrated, establishing the model upon which the next iterative formula is ascertained. The two outermost adjacent pairs were inspected, with the larger of the two being selected as $n_3$ (orange). The assignment of $n_2$ and $n_1$, indicated as light blue and green arrows, respectively, followed an alternating succession. (b) Once the $n_j$ terms were assigned, the CER was calculated as a function of $n_3$ based on a first prediction inspired by Eq. (2), shown as a red curve that did not match the corresponding simulations (black squares). The introduction of a corrective term, $\delta_3$, restored agreement between the formula and the simulations.

### 4.3. Adjacent Terminals of Similar Polarity and the Unit Terminal

Before jumping to the generalized case for any number of terminals, describing other configurations that do not alternate in polarity will be most beneficial as a tool to simplifying many complex configurations. For more information, please consult the Supplemental Material [51]. In Sec. IV A, the edge state picture and Fig. 3 were presented under the condition that a term $n_x$ was used to describe the junctions between two adjacent terminals of similar polarity. Since treating a self-interacting, non-alternating terminal configuration did not obey the same formula as its alternating counterpart, using $n_x$ was the simplest way to characterize the configuration. A similar configuration to Fig. 3 was used in Fig. 5 (a), and prior to its analysis, self-interaction was verified with the edge state picture. The following formula was used as an ansatz with the intention of analyzing the limits of $n_x$:

$$q_3(n_x) = \frac{xn_x + x}{yn_x + z}$$

(4)

Equation (4) (which for this configuration is only valid for even $n_x$) takes on the form of Eq. (3), with the numerator fully distributed and the denominator containing two terms. As seen in Fig. 5 (a), if the limits of $n_x$ are evaluated, these variables can be found:

$$\lim_{n_x \to \infty} \frac{xn_x + x}{yn_x + z} = \frac{x}{y}, \text{and} \lim_{n_x \to 0} \frac{xn_x + x}{yn_x + z} = \frac{x}{z}$$

(5)

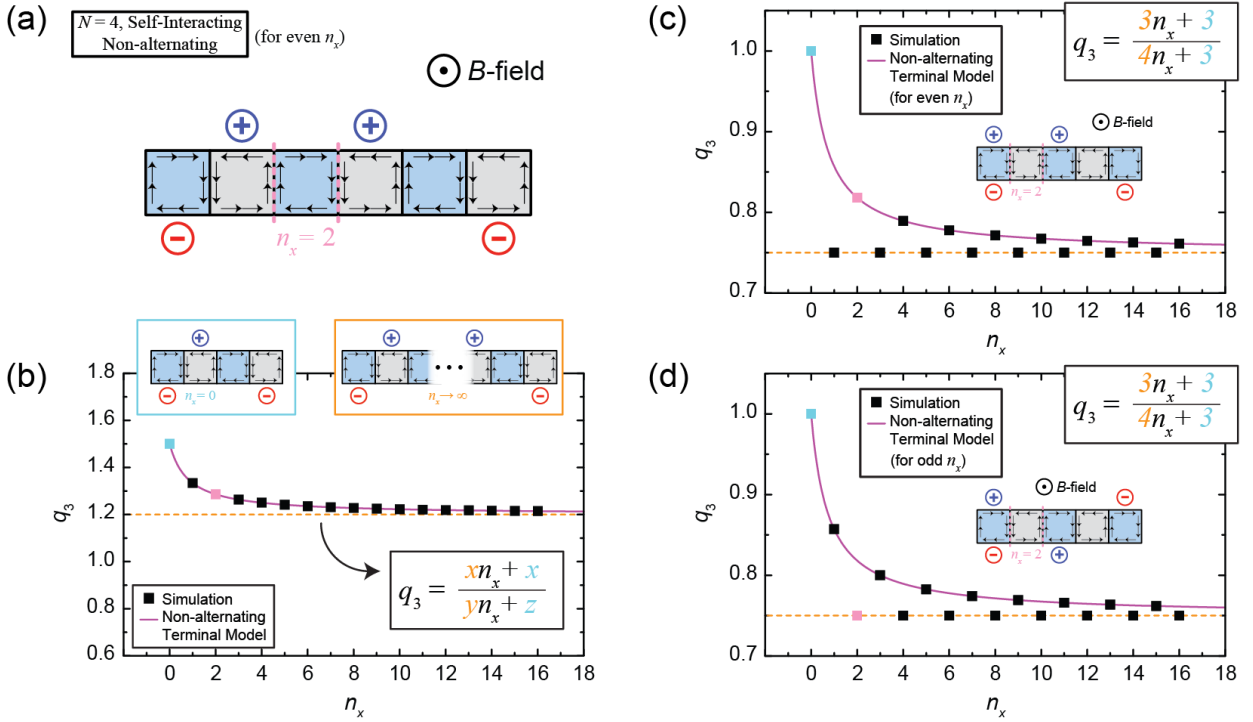

FIG. 5. (Color online) (a) A self-interacting configuration is shown with a non-alternating sequence of terminals (going from left to right). Treated differently than its alternating counterpart, these configurations are best understood by the limiting behavior of the number of junctions between the two like-polar terminals, labeled as $n_x$ in pink. (b) Data from the LTspice simulations are shown as squares. The two limiting cases are shown in light blue and orange for $n_x \to 0$ and $n_x \to \infty$, respectively. The CER in those limits are also color coded based on the corresponding limit. The limited-case CERs are used to construct the final formula valid for all $n_x$. (c) and (d) Using unit terminals, the symmetry of odd and even cases of $n_x$ is shown. When part of the configuration is reflected about the horizontal, the edge state picture dictates that the configuration will switch its self-interaction condition (so if it starts as self-interacting, then it will switch to having isolated components).

Equation (5) yields two ratios that need to be determined to obtain a formula for this configuration. For $n_x \to \infty$, the left and right pairs of terminals are considered isolated, thus equating $\frac{x}{y}$ to the CER as if both pairs were connected in parallel, and this value is $\left(\frac{1}{2} + \frac{1}{3}\right)^{-1} = \frac{6}{5}$.

In Fig. 5 (b), the isolated-case CER manifests itself as the asymptotic value for $q_3$ (dotted orange line), with its hypothetical device depicted in the orange outlined box. For $n_x = 0$, the whole configuration effectively becomes the one shown in the light blue outlined box, which is an alternating terminal configuration that obeys Eq. (2). The CER for this limiting case is $q_2 = \frac{3}{2}$, and this implies that, by taking the lowest common factors, $x = 6$, $y = 5$, and $z = 4$, the final expression for this configuration for all $n_x$ is $q_3(n_x) = \frac{6n_x + 6}{5n_x + 4}$. Ergo, for the exact configuration in Fig. 5 (a), shown as the pink data point in (b), $q_3 = \frac{9}{7}$.

The final two conditions to mention are those of the "unit" terminal, in which a positive and negative terminal both occupy the same region, and the existing symmetry of the formulas for non-alternating terminal configurations. First, the unit terminal has the obvious CER value of 1 when isolated. Once incorporated into a larger configuration, like the one shown in Fig. 5 (c), the edge state picture is needed to determine whether the unit terminal will remain isolated or interact with the other terminals. For the graph in Fig. 5 (c), full self-interaction is expected only for even $n_x$, and for all odd $n_x$, the CER is $\frac{3}{4}$. The unit terminal immediately renders the configuration as one that is non-alternating, warranting the approach taken for Eq.s (4) and (5). By repeating the analysis, one arrives at $q_3(n_x) = \frac{3n_x + 3}{4n_x + 3}$ for even $n_x$ (but for cases when the unit terminal is near another single terminal of either polarity, the total CER remains unaffected and at 1 – see the Supplemental Material [51]). With the formula for even $n_x$, one finds $q_3 = \frac{9}{11}$ (pink data point in Fig. 5 (c)).

To address the second condition, if the two, adjacent, like-polar terminals are on opposite sides of the device, as shown in Fig. 5 (d), then this reflection about the horizontal must then be described by the same equations, with the conditions on even and odd $n_x$ being switched. The repeated simulations and mathematical analysis are plotted, and as expected, the CER, still being based on $n_x = 2$, now describes the parallel combination of two isolated configurations ($q_3 = \frac{3}{4}$).

### 4.4. The Generalized Case

To begin the process of generalizing this behavior for alternating configurations with any *N*, it will help to go through one additional analysis of limiting cases. Recall Eq. (3), with appropriate subscripts for the arbitrary *N* = 5 case:

$$q_4(n_4) = \frac{q_3(n_4 + 1)}{q_3 + n_4 + \delta_4}$$

$$(6)$$

One note to make is that $\delta_4$ does not depend on $n_4$, but rather the remaining $n_j$ terms, and that independence allows for an easier inspection of the limiting case when $n_4 = 0$. By applying this case to Eq. (6), the relation for the corrective term becomes clear: $\delta_4 = q_3\left(\frac{1}{q_4^{(0)}} - 1\right)$. By substituting this back into Eq. (6), the following result is obtained:

$$q_4(n_4) = \frac{q_3(n_4 + 1)}{n_4 + \frac{q_3}{q_4^{(0)}}}$$

$$(7)$$

In Eq. (7), the term $q_4^{(0)}$ is a CER that characterizes a similar *N* = 5 configuration as above, with the substantial difference that one of its outermost adjacent terminal pairs (parameterized by $n_4$) has been reduced to a unit terminal. As learned in Sec. IV C, a unit terminal renders a configuration *non-alternating*, and so a new calculation must be performed to obtain $q_4^{(0)}$. After that calculation or simulation, $q_4(n_4)$ is fully known and can be predicted (an example is provided in the Supplemental Material [51]). The final result is one where Eq. (7) has been generalized for any alternating configurations containing $N$ terminals:

$$q_{N-1}(n_{N-1}) = \frac{q_{N-2}(n_{N-1} + 1)}{n_{N-1} + \frac{q_{N-2}}{q_{N-1}^{(0)}}}$$

(8)

Combining all the presented mathematical techniques enables one to predict the CER for an arbitrary configuration of large $N$ despite the increasing complexity of both the configurations and the corresponding calculations.

## V. EXPERIMENTAL SUPPORT

Two functioning devices were fabricated to test this theoretical framework. They were inherently limited by the number of allowable electrical contacts and the good performance of a subset of those contacts. Nonetheless, experimental data were acquired at ± 9 T, 1.6 K, and on a transistor outline (TO-8) package. Two relatively complex configurations that fit within the useable contacts were measured and plotted in Fig. 6. The first configuration was a non-alternating one that was not self-interacting (since $n_x = 1$). It essentially contained a unit terminal on one isolated branch and a CER of 2 on the other. A quick parallel resistance calculation would reveal an expected value of $q_3 = \frac{2}{3}$, and this comparison is made on Fig. 6 (a) between the purple data and dashed orange line for the theory.

Fig. 6 (b) has an additional terminal, but because $n_x = 1$ (odd) and the two like-polar terminals are on the same side of the device, the unit terminal on the left is still isolated from the rest of the configuration. The right side of $n_x$ may be treated as an isolated $N = 3$ configuration (with subsequent assignments of $n_1$ and $n_2$). The right side has a CER of $\frac{3}{4}$ per Eq. (2), and when this value is placed in parallel with the left side's CER of 1, the expected value is $q_4 = \frac{4}{7}$. In both cases, there is a small offset from the plateaus, most likely due to instrument errors and device imperfections. More experimental efforts are called for given the extensive parameter space that can be explored. Any simulations performed henceforth may now be trusted and understood, seeing as its analytical counterpart has been explained and expressed.

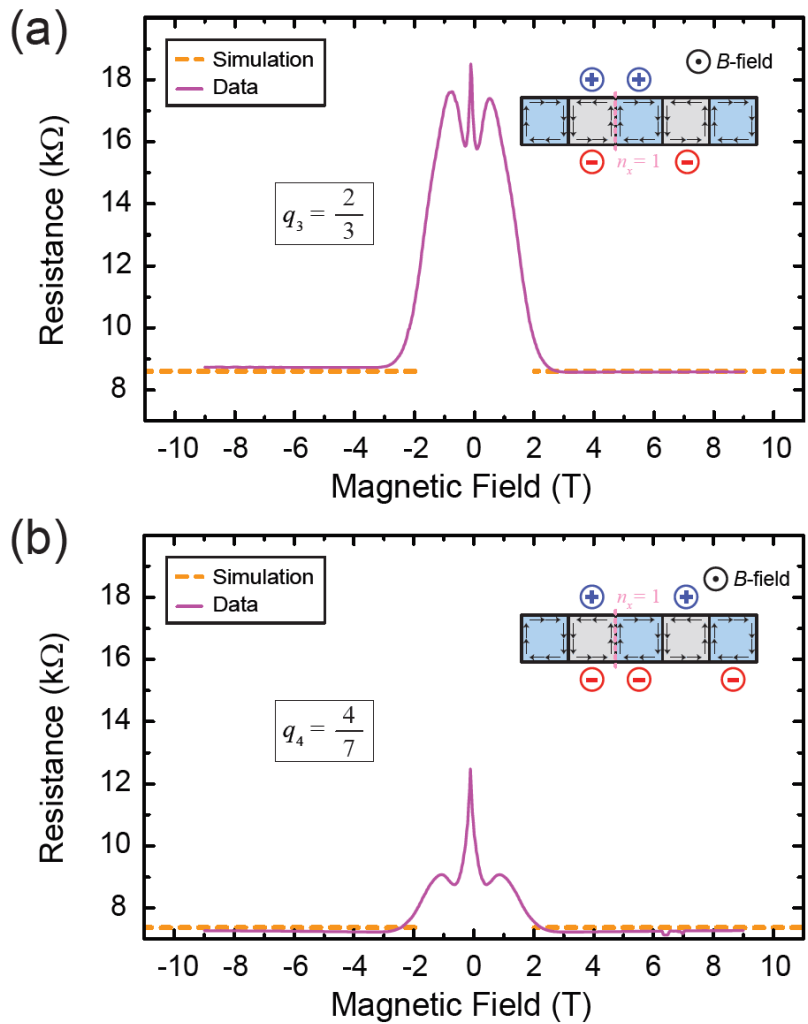

FIG. 6. (Color online) (a) Experimental data are shown in purple for a non-alternating, non-self-interacting, $N = 4$ configuration. The configuration is essentially two isolated $N = 2$

configurations, one of them being a unit terminal. The calculation and simulation are together represented by the dotted orange line at the resistance value of $\frac{2}{3} R_{\text{H}}$. (b) Experimental data are shown in purple for a non-alternating, non-self-interacting, $N = 5$ configuration. The configuration is essentially two isolated configurations - one ($N = 2$) unit terminal and one ($N = 3$) alternating, self-interacting terminal. The calculation and simulation are together represented by the dotted orange line at the resistance value of $\frac{4}{7} R_{\text{H}}$.

In Ref. [49], similar measurements were performed without the knowledge of how the obtained CERs came about. By applying Eq. (8) and the preceding conditions, the same CERs were predicted and match the simulations and their corresponding experimental results.

## VI. CONCLUSIONS

In conclusion, a new mathematical approach to predicting quantized resistances in *pn*J devices was introduced, and it incorporated multiple entry and exit points for electron flow. This incorporation allowed for the mathematical prediction of the nonintuitive, effective resistances resulting from the redistribution of the electric potential in the devices. The theoretical framework was supported by experimental data taken exclusively at the $\nu = 2$ plateau, confirming the ability to measure fractional multiples of $R_{\text{H}}$. A more fundamental consequence of the resulting generalized framework is that, for a number of applications that require quantized resistances, measured values can essentially be customized, thus opening a whole new avenue of exploration for quantum Hall efforts.

### ACKNOWLEDGMENTS

The work of MM at NIST was made possible by M Ortolano of Politecnico di Torino and L Callegaro of Istituto Nazionale di Ricerca Metrologica, and the authors thank them for this endeavor. The work of DKP at NIST was made possible by C-T Liang of National Taiwan

University, and the authors thank him for this endeavor. The authors would like to express thanks to L Chao and S Le for their assistance in the NIST internal review process. The authors declare no competing interests.

United States government, nor is it intended to imply that the materials or equipment identified are necessarily the best available for the purpose.

GRAPHICAL ABSTRACT (END OF DOC)

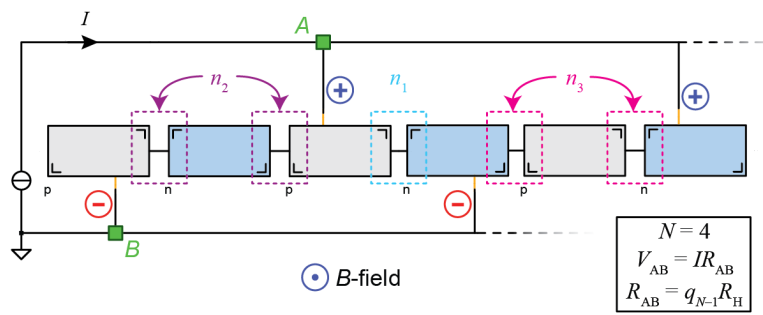

Supplemental Material

Algorithm for constructing customized quantized resistances in graphene *p-n* junctions

Albert F. Rigosi[1][†], Martina Marzano[2,3], Antonio Levy[1], Heather M. Hill[1], Dinesh K. Patel[1,4], Mattias Kruskopf[1,5], Hanbyul Jin[1,5], Randolph E. Elmquist[1], and David B. Newell[1]

[1]*Physical Measurement Laboratory, National Institute of Standards and Technology (NIST), Gaithersburg, MD 20899, United States*

[2]*Department of Electronics and Telecommunications, Politecnico di Torino, Torino 10129, Italy*

[3]*Istituto Nazionale di Ricerca Metrologica, Torino 10135, Italy*

[4]*Department of Physics, National Taiwan University, Taipei 10617, Taiwan*

[5]*Joint Quantum Institute, University of Maryland, College Park, MD 20742, United States*

Contents

[†] Email: afr1@nist.gov

# 1. Examples of Complex Configurations

An example of one large, complex configuration, along with some of its sub-configurations, will be provided. The main configuration ($N = 11$) is shown in Fig. 1-SM. The calculations will be done step-by-step for the first few components of the configuration, after which the remainder will be left as an exercise to the reader.

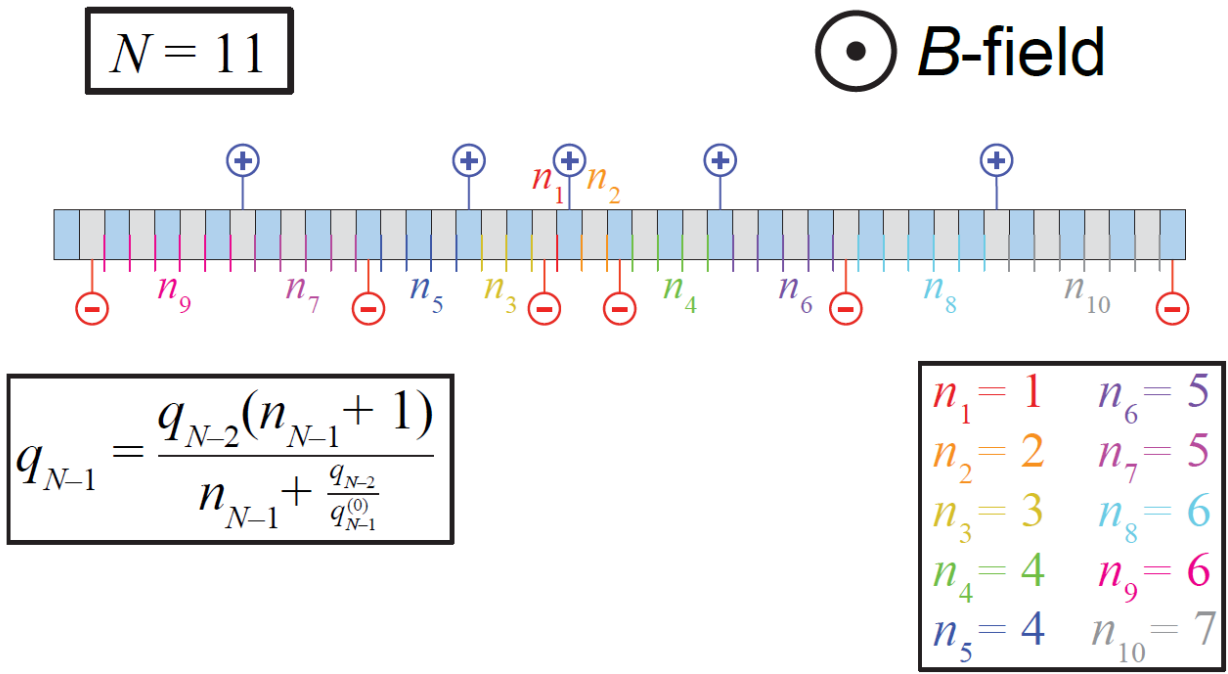

**FIG. 1-SM.** (Color online) A complex configuration is provided as the first example of applying the theoretical framework presented in the main text. This $N = 11$ alternating configuration has its $n_j$ terms assigned starting with the larger of the two outermost pairs of adjacent terminals.

Once the $n_j$ terms have been assigned, the calculation may begin. As described in the main text, selection of the larger of the two outermost pairs of adjacent terminals to be $n_{10}$. When assigning the lower numerical subscripts, take care to go from one side of the device to the other.

This will result in the even subscripts being on one side and the odd subscripts being on the other. This condition is necessary to ensure the exactness of Eq. (A1) (also in Fig. 1-SM):

$$q_{N-1}(n_{N-1}) = \frac{q_{N-2}(n_{N-1} + 1)}{n_{N-1} + \frac{q_{N-2}}{q_{N-1}^{(0)}}}$$

(A1)

With $n_1 = 1$, we use the linear form of Eq. (A1), which is $q_2 = (n_1 + 1)$. Note that this is the only case when Eq. (A1) "breaks down", mainly because of the ill-defined $q_0$. $q_1 = 2$, and that brings us to $n_2 = 2$. If we apply Eq. (A1) again:

$$q_2(n_2) = \frac{q_1(n_2 + 1)}{n_2 + \frac{q_1}{q_2^{(0)}}} = \frac{2(2 + 1)}{2 + \frac{2}{1}} = \frac{6}{4} = \frac{3}{2}$$

In the main text, the term $q_2^{(0)}$ was not directly addressed since the simpler configurations preceded the definition of that term. Please consult the third section for details on why this term is 1.

$$q_3(n_3) = \frac{q_2(n_3 + 1)}{n_3 + \frac{q_2}{q_3^{(0)}}} = \frac{\frac{3}{2}(3 + 1)}{3 + \frac{3/2}{3/4}} = \frac{6}{5}$$

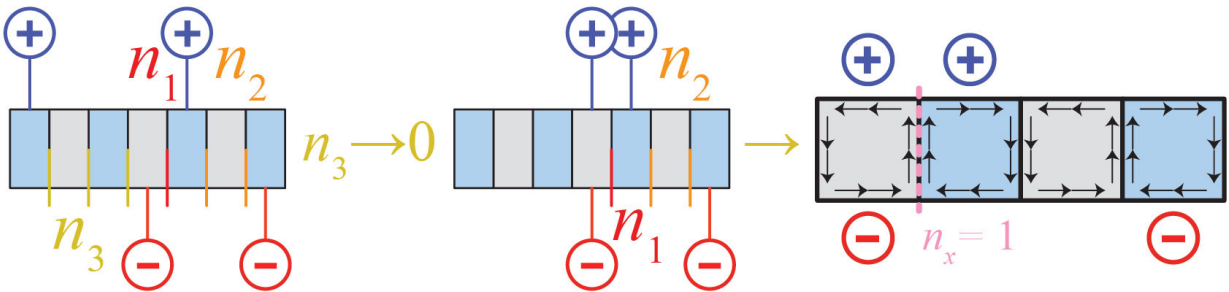

**FIG. 2-SM.** (Color online) A sub-calculation of the complex configuration from Fig. 1-SM is visualized. To solve for the $q_3^{(0)}$ term, which is the case where $n_3 \to 0$, the new configuration must have its total CER calculated. The edge state picture will help us determine if there are components that are isolated from one another.

To calculate the $q_3^{(0)}$ term, $n_3 \to 0$, and resulting configuration must be solved as though it was a new configuration. This resembles the configuration in Fig. 6 (a) of the main text. The edge state picture suggests that the electron flow from each of the two negative terminals cross in opposite directions at the single $n_x$ junction. Please consult the fourth section for why this opposite direction flow results in cancellation. Now we know that this new $q_3^{(0)}$ configuration in Fig. 3-SM is composed of two isolated branches. The left branch (a unit terminal) has a CER of 1, and the right branch has a CER of 3. Taking the parallel addition of the two yields $q_3^{(0)} = \frac{3}{4}$. Using this value gives $q_3 = \frac{6}{5}$.

$$q_4(n_4) = \frac{q_3(n_4 + 1)}{n_4 + \frac{q_3}{q_4^{(0)}}} = \frac{\frac{6}{5}(4 + 1)}{4 + \frac{6/5}{2/3}} = \frac{30}{29}$$

Moving on to $q_4$, we repeat the calculation, knowing all the terms except for $q_4^{(0)}$. To get this value, we must repeat the previous analysis of taking the limiting case of $n_4 \to 0$, giving us $q_4^{(0)} = q_4(n_4 = 0)$. The relevant portion of the configuration from Fig. 1-SM is used for Fig. 3-SM below to enhance the clarity.

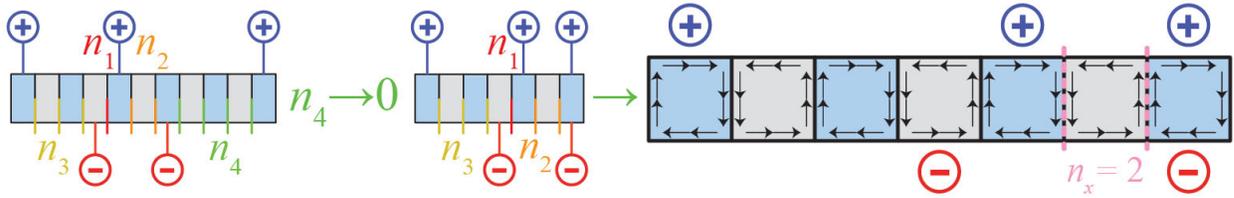

**FIG. 3-SM.** (Color online) A sub-calculation of the complex configuration from Fig. 1-SM is shown. To solve for the $q_4^{(0)}$ term, which is the case where $n_4 \to 0$, the new configuration must have its total CER calculated. The edge state picture will help us determine if there are components that are isolated from one another.

The corresponding configuration for $q_4^{(0)}$ is on the right side of the green arrow in Fig. 3-SM. The edge state picture suggests that the unit terminal on the right interacts with the other terminals, making this configuration self-interacting (see Fig. 3 of the main text). We use the ansatz in Eq. (4) of the main text:

$$q_4^{(0)}(n_x) = \frac{x n_x + x}{y n_x + z}$$

Consider the case when $n_x \to \infty$. The ratio $\frac{x}{y}$ is equal to the total CER of two independent branches in parallel, with the left branch having a CER of $\frac{8}{5}$ (in this new configuration's left branch, $n_1 = 1$ and $n_2 = 3$) and the right branch having a CER of 1 (unit terminal). Therefore, $\frac{x}{y} = \frac{8}{13}$. Now let's consider the $n_x \to 0$ case. The ratio $\frac{x}{z}$ is equal to the CER of the configuration shown in Fig. 4-SM.

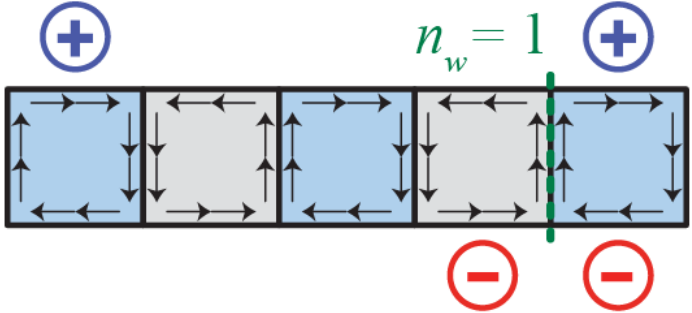

**FIG. 4-SM.** (Color online) A sub-sub-calculation of the complex configuration from Fig. 3-SM is provided. Every additional terminal increases the complexity of the calculation since it will typically require sub-calculations of slightly simpler configurations. The new term $n_w$ is used as we calculate the CER of the depicted configuration.

To make matters slightly easier, the CER of Fig. 4-SM is the parallel combination of $n_{\text{left}} = 4$ and $n_{\text{right}} = 1$, giving $\frac{x}{z} = \frac{4}{5} = \frac{8}{10}$. Now we may return to the final steps of determining $q_4^{(0)}$. Plugging in the lowest common factors for the ratios yields:

$$q_4^{(0)}(n_x = 2) = \frac{xn_x + x}{yn_x + z} = \frac{8n_x + 8}{13n_x + 10} = \frac{24}{36} = \frac{2}{3}$$

Now that $q_4(n_4 = 4) = \frac{30}{29}$ has been established, the calculation may continue with $n_5 = 4$. At this point, the remaining calculations are left as an exercise to the reader, and they will find that the simulations match the resulting calculated values:

$$q_5(n_5 = 4) = \frac{15}{17}, q_5^{(0)} = \frac{5}{9}$$

$$q_6(n_6 = 5) = \frac{15}{19}, q_6^{(0)} = \frac{15}{29}$$

$$q_7(n_7 = 5) = \frac{5}{7}, q_7^{(0)} = \frac{15}{31}$$

$$q_8(n_8 = 6) = \frac{21}{32}, q_8^{(0)} = \frac{15}{34}$$

$$q_9(n_9 = 6) = \frac{105}{173}, q_9^{(0)} = \frac{105}{251}$$

$$q_{10}(n_{10} = 7) = \frac{420}{737}, q_5^{(0)} = \frac{105}{263}$$

Having such a device at the $\nu = 2$ plateau ($R_H \approx 12906\ \Omega$) would result in a quantized effective resistance of: $R_{AB} = q_{10}R_H \approx 5152.75\ \Omega$.

## 2. Magnetic Field Symmetry Condition

This section is meant to support the assumption that terminal configurations yield the same CER for both polarities of magnetic field. The support will stem mostly from a visual representation of the edge state picture. In Fig. 5-SM, non-alternating configurations are placed in both polarities of magnetic field. A self-interacting configuration Is shown in Fig. 5-SM (a) and (b), with positive and negative $B$-field, respectively. The difference in how the currents are permitted to travel does not alter the configuration's status as one that is self-interacting.

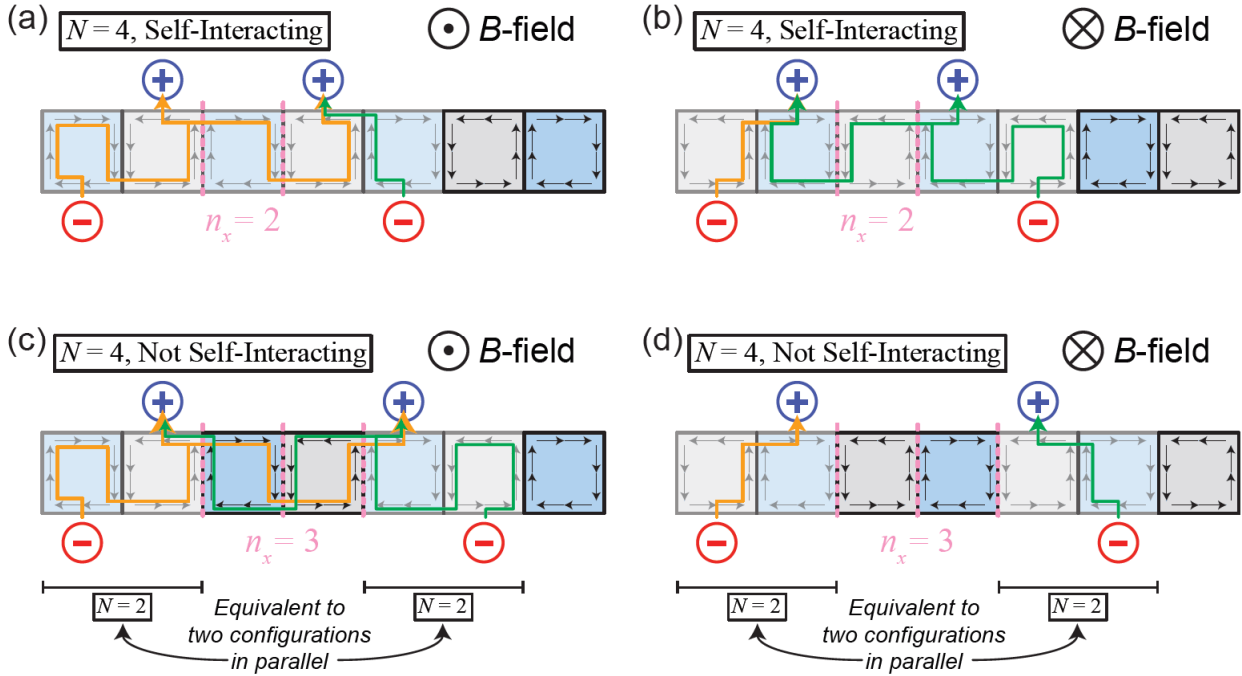

**FIG. 5-SM.** (Color online) A self-interacting, non-alternating terminal is shown at (a) positive and (b) negative $B$-fields. The difference in electron flow does not change the configuration's status, and simulations show that the CER remains the same. (c) and (d) A similar set of configurations is shown that is not self-interacting. In both cases, the isolation of both branches occurs by slightly different ways.

A configuration that is not self-interacting is shown in Fig. 5-SM (c) and (d), with positive and negative $B$-field, respectively. Though the differences in electron flow are more substantial, both cases still feature central opaque regions that effectively isolate the two pairs of terminals. The mechanism for edge state "cancellation" in Fig. 5-SM (c) is shown in section four below. One final note to make is that the cases of alternating configurations are easier to justify as having this CER symmetry with positive and negative $B$-field because no matter which region electrons begin to flow from, the first junction will split the current such that they flow towards both positive, adjacent terminals.

### 3. Unit Terminal and Adjacent, Like-Polar Terminals on Opposite Device Sides

To elaborate on an earlier detail in section one, $q_2^{(0)} = 1$ was a result used for the example calculations above. If one considers where a unit terminal has a third terminal adjacent to it (see Fig. 6-SM (a)), we can arrive at that result by performing the limiting case analysis.

$$q_3(n_x) = \frac{xn_x + x}{yn_x + z}$$

If $n_x \to 0$, the ratio $\frac{x}{z}$ is 1 since all that remains is a single unit terminal. When $n_x \to \infty$, the lone negative terminal at infinity will provide no current due to the infinite resistance that path would offer to an electron. Ergo, the ratio $\frac{x}{y} = 1$, which is the case where current only flows through the unit terminal. It is implied that $y = z$, and the final result for this configuration is that $q_3^{(a)}(n_x) = 1$, independent of $n_x$. Since Fig. 6-SM (a) and (b) are both non-alternating cases with the two negative terminals separated by $n_x$, a similar result will hold true: $q_3^{(b)}(n_x) = 2$.

Next, we present another example of like-polar terminals that are either on the same side of the configuration or on opposite sides. The former case is presented in Fig. 6-SM (c), where if $n_x \to 0$, the ratio $\frac{x}{z}$ is 1 again. When $n_x \to \infty$, the ratio $\frac{x}{y}$ is equal to the CER of two parallel unit terminals. Therefore, we get $\frac{x}{y} = \frac{1}{2}$, and the two ratios give us our final result for even $n_x$:

$$q_2(n_x) = \frac{n_x + 1}{2n_x + 1}$$

When the like-polar terminals are on opposite sides, the formula remains true for odd $n_x$.

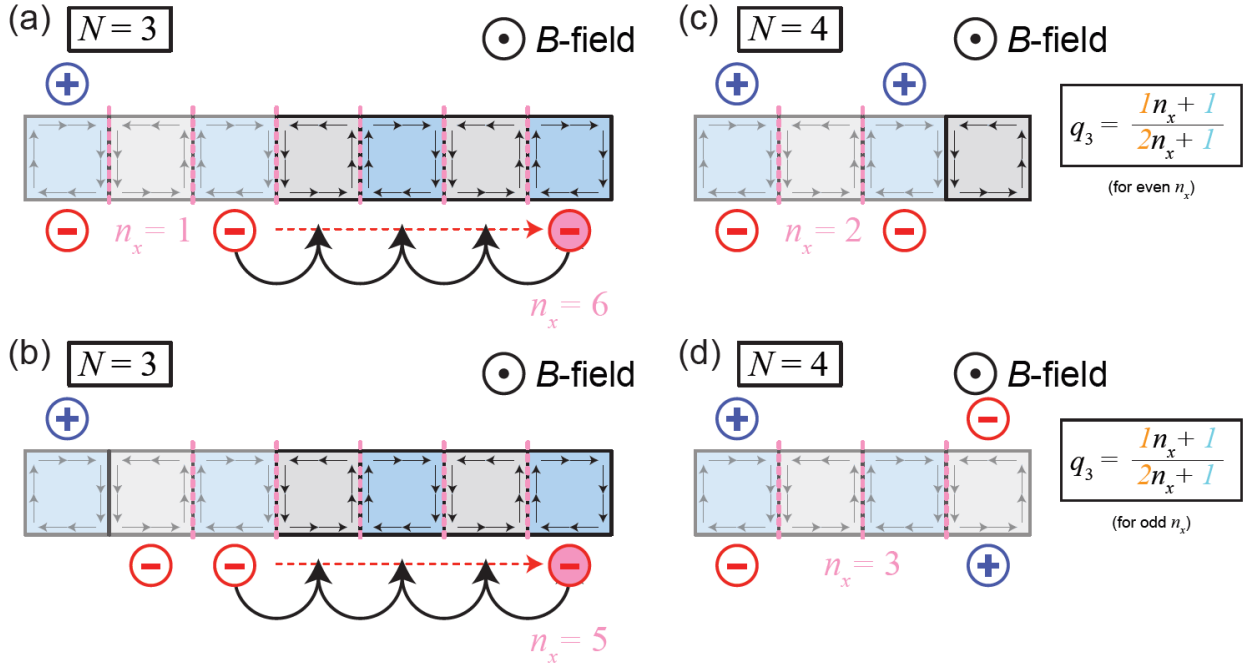

**FIG. 6-SM.** (Color online) (a) The CER for a unit terminal remains unaffected when a third terminal is introduced, and the formula simplifies to a constant function of $n_x$. (b) The same result as (a) can be seen for a configuration that has two like-polar terminals at its endpoint. (c) and (d) The even and odd application of a formula for $n_x$ is shown to switch depending on the orientation of the two like-polar, adjacent terminals (and in this special case where the configuration is doubly non-alternating, it does not matter whether we focus on the positive terminals or negative terminals, for both pairs are at opposite sides of one another).

### 4. Equilibration in Non-Self-Interacting Configurations

One detail that needs to be clarified is the visual representation of the case when portions of a configuration are isolated from one another. As shown in other figures, there are two methods by which a configuration can have isolated components. The first, and more straightforward of the two, is when electrons are unable to flow through portions of the configuration because the electric potential extremum has been reached, i.e. the electron has reached the positive terminal along its path leading out of the circuit (Fig. 5-SM (d)).

The second case is one that warrants a more in-depth explanation. Fig. 7-SM is a magnification of part of Fig. 5-SM (c) (and essentially part of Fig. 3 in the main text). If one current is larger than the other, at every junction crossing point, then a portion of the larger current will split off to return toward the positive terminal it had already passed. This further reduction of the larger current continues to the point that both currents within the middle regions are equal at the junction crossing points. Because the are equal and opposite, the net current flow is zero, and those central regions do not contribute to the CER. This situation is equivalent to having the two branches in the circuit in parallel.

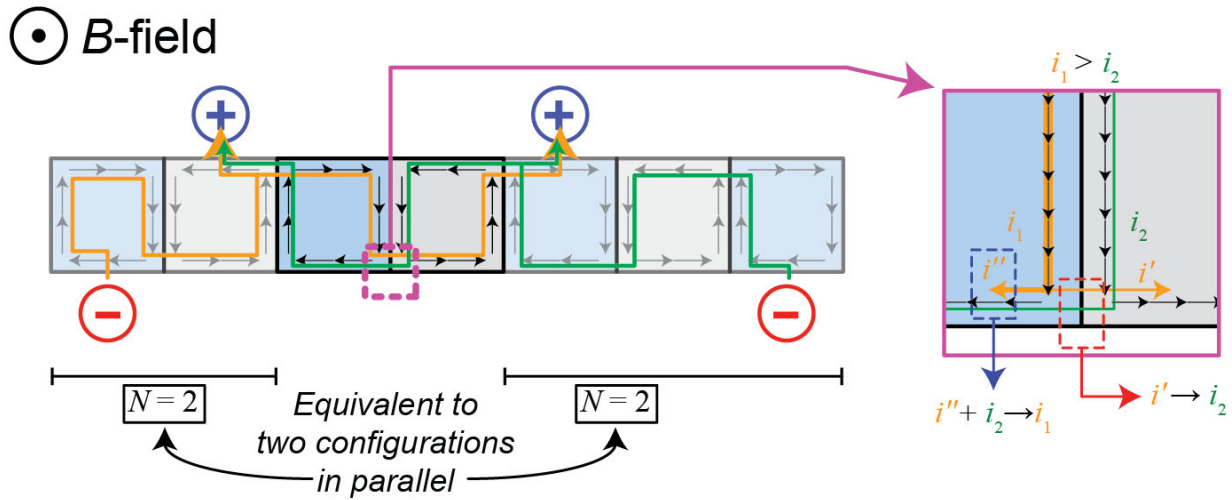

**FIG. 7-SM.** (Color online) A semiclassical visualization of the edge currents canceling at junction crossing points is illustrated. Simply put, in cases where $i_1 > i_2$, a portion of $i_1$ (orange paths) will not cross the junction ($i''$) due to the difference in electric potential. The portion that crosses the junction, $i'$, will approach the value of $i_2$, and in the steady-state, the two values will match.

## 5. Zero-Field Graphene Response Data

To give an idea of the electrical response of graphene at zero-field, three sets of measurements were taken on the completed device as it was undergoing exposure to ultraviolet light. This exposure was modulating the doping in the two *p*-type regions while keeping the *n*-type region mostly unchanged. Photochemical gating is demonstrated here as one method of creating these large-scale devices (see Fig. 8-SM).

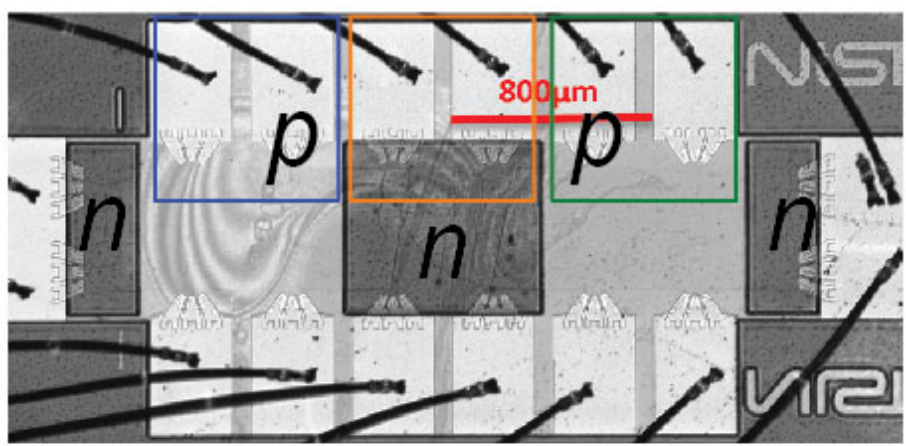

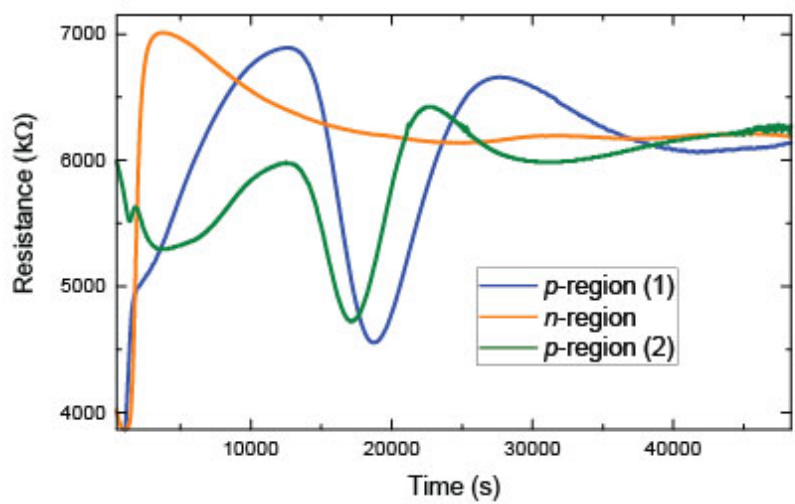

**FIG. 8-SM.** (Color online) An image of an experimental device used for the magnetoresistance measurements in the main text. At zero-field, the photochemical gating response is recorded as a function of time for three pairs of longitudinal resistances.